# Nuclear medium effects on the properties of $\Lambda(1405)$

**J. Y. Süngü**

Department of Physics, Kocaeli University, 41001, İzmit, Türkiye

**N. Er**

Department of Physics, Bolu Abant İzzet Baysal University, Gölköy Kampüsü, 14030, Bolu, Türkiye

E-mail: `jyilmazkaya@kocaeli.edu.tr, nuray@ibu.edu.tr`

**Abstract.** We study the $\Lambda(1405)$ resonance with $I(J^P) = 0(1/2^-)$ in the context of the pentaquark hypothesis in nuclear medium. For the investigation of the influence of the nuclear medium on the physical parameters of $\Lambda(1405)$, we propose a molecular-type structure involving $K^- p$ and $\bar{K}^0 n$ admixtures, correlated with the nuclear matter density. Our analysis reveals a substantial shift in both mass and residue, approximately 20% and 38%, respectively. These findings have significant implications for experimental researchers aimed at identifying in-medium characteristics of hyperon resonances.

## 1. Introduction

Exploring the nature of the $\Lambda(1405)$ resonance remains a persistent and complex topic of research initiated since its discovery in 1961 [1]. Three-quark models generally coincide well with experimental outcomes for numerous hadronic states. Nonetheless, the $\Lambda(1405)$ poses a challenge to the conventional framework. Because, even though it contains strange quark in its composition, it has lighter mass than its non-strange counterpart, $N(1535)$. Additionally, the observed mass difference from the $\Lambda(1520)$ state, assuming the spin-orbit partner of $\Lambda(1405)$, stands out as significantly larger compared to the nucleon sector. The mass distribution deviates from a Breit-Wigner pattern. While the Breit–Wigner widths of $\Lambda(1520)$ and $\Lambda(1405)$ are 16 and 50 MeV, the corresponding states $N(1520)$ and $N(1535)$ exhibit widths of 110 and 150 MeV, respectively. Furthermore, investigation into $N_c$ scaling reveals that the existence of the three-quark configuration in the $\Lambda(1405)$ structure is unlikely. Remarkably, the electromagnetic size of $\Lambda(1405)$ is measured to be larger than that of typical three-quark hadrons. The difficulties of the three-quark picture led Dalitz and Tuan [2] to predict it as a quasi-bound state of $\bar{K}p$ encapsulated in the $\Sigma\pi$ continuum. Analysing the observed mass spectrum of $\pi^-\Sigma^+$, Hemingway reported a resonance mass of $(1406.5 \pm 4.0)$ MeV and a width of $(50 \pm 2)$ MeV for $\Lambda(1405)$. This assessment was conducted using $\bar{K}N$ scattering theory, as detailed in the studies by Dalitz and Tuan [3,4]. The Particle Data Group (PDG) [5] presents average mass and width values of $\Lambda(1405)$ resonance as:

$$M_\Lambda = 1405.1^{+1.3}_{-0.9} \text{ MeV},$$
$$\Gamma_\Lambda = 50.5 \pm 2.0 \text{ MeV}. \tag{1}$$

Furthermore, an alternative framework proposing a double-pole structure for $\Lambda(1405)$ has emerged based on chiral $SU(3)$ dynamics. In this model, two poles for $\Lambda(1405)$ are suggested, positioned between 1420 and 1390 MeV/c$^2$, intricately linked with the $\bar{K}N$ and $\Sigma\pi$ channels, respectively [6–8]. The $\Lambda(1405)$ should be a mixture of these two states because all possible states with the same quantum number can combine via the strong force. To understand the inner structure of the $\Lambda(1405)$ resonance and gain insights into $\bar{K}N$ dynamics, it is essential to determine the dominant component, as discussed in [9]. Recently there have been some theoretical studies for $\Lambda(1405)$ assuming a molecule-like structure of $\bar{K}N$ [10–14]. It is probably the first exotic state to have a meson-baryon or pentaquark structure. The latest Particle Data Group review [5] cautiously acknowledges the potential presence of an additional resonance pole for $\Lambda(1380)$ with limited confidence. While recent studies suggest the $\Lambda(1405)$ particle might have two "poles" within its energy existence, and a narrow resonance seems evident just under the interaction threshold of kaons and nucleons, the properties and exact location of a potential second resonance remain unclear. The double-pole structure of $\Lambda(1405)$ makes this state particularly attractive, while we know now that such a structure is common to many other states in the hadron spectrum.

The interesting presence of the $\Lambda(1405)$ state below the $\bar{K}N$ threshold adds a distinctive aspect to the $\bar{K}N$ interaction potential in the nuclear medium. Phenomenological potential models treat the $\Lambda(1405)$ as a $\bar{K}N$ bound state with a binding energy of 27 MeV [15, 16]. Lattice QCD calculations give more support for the molecule-like structure against to ordinary three-quark state [17, 18]. Ref. [19] used a chiral unitarity model, where the resonance is generated dynamically from $\bar{K}N$ interactions with other channels produced from the octets of baryons and mesons, shows that the line shape would rely on the decay mode. The assessments indicate variations in the line shape for different decay modes, and the measurements for charged decays deviate from the predictions. Moreover, in the work conducted by Ref. [20], a comprehensive analysis of the $\bar{K}N$ two-body scattering amplitude in the $I = 0$ channel near the $\bar{K}N$ mass threshold reveals the extraction of a resonance pole at $1417.7^{+6.0}_{-7.4}(fit)^{+1.1}_{1.0}(syst.)$MeV/c$^2$. While a thorough analysis presented in Ref. [21] leans towards a single resonance, it does not conclusively dismiss the possibility of a two-pole scenario.

The strangeness sector of hadrons consists of some unusual resonances that indicate unexpected properties. Among them, only a few other states have such a long history as the mysterious $\Lambda(1405)$ [22]. Many experimental data on resonance $\Lambda(1405)$ have been announced [23–28]. This state manifests itself in the meson-baryon scattering amplitude with a spin-parity of $J^P = (\frac{1}{2})^-$, strangeness $S = -1$, and isospin $I = 0$. In 2011, the adoption of novel triggerable X-ray detectors developed within the SIDDHARTA project facilitated the most precise measurement of kaonic hydrogen K-series X-rays, leading to a significantly improved settling of the $\bar{K}p$ scattering length [23]. The spin-parity of the $\Lambda(1405)$ is determined with $J^P = 1/2^-$ using photoproduction data in the decay channel $\Lambda(1405) \rightarrow \Sigma^+ + \pi^-$ from the CLAS detector at Jefferson Lab [26]. The $\bar{K}N$ interaction's coupled-channel nature was verified through the exploration of $\bar{K}p$ correlations in proton-proton collisions at the LHC, as measured by ALICE [29]. Although the ALICE [27] and BGOOD data [28] confirm two-pole assumption, J-PARC [20] claim a single pole, closer to the $\bar{K}N$ mass threshold than the value given by the PDG. Recently, Anisovich et al. presented a single-pole representation for the $\Lambda(1405)$ contribution, fitting the data from Ref. [28] in $\gamma$ and $K^-$ induced reactions on protons and the kaonic hydrogen atom. They reported a pole position of $(1422 \pm 3)$ MeV/c$^2$ [30].

Research in the strangeness of nuclear physics traces back to the detection of the first hyperon, the $\Lambda$, in cosmic rays. The behavior of a $\Lambda$ within a nuclear medium is primarily influenced by hadronic interactions among baryons and presents a complex many-body problem. This makes the $\Lambda$ a valuable probe for understanding nuclear features that are not easily elucidated through other techniques. So, the $\Lambda$ is a prominent baryon within the nucleus [31]. Describing the $\Lambda(1405)$ as a $K^-p$ bound state, Koch et al. obtain that the mass of the $\Lambda(1405)$ is relocated upwards in energy above the $K^-p$ threshold due to the Pauli blocking of intermediate states [32]. The $\Lambda(1405)$ resonance experiences very strong variations in the nuclear medium at densities nearly $(0.1-0.2)\rho_0$ dissolving chiefly through Pauli blocking of the intermediate proton states [33–35]. At nuclear saturation density, it is predicted attractive mass shifts for the Λ(1405) nearly 60 MeV [36].

Moreover, as indicated in Ref. [37], the Λ(1405) emerges as the dominant exotic particle within stars, exhibiting prominence at different entropies for both neutrino-free and neutrino-trapped stellar matter. This observation stems from a relativistic model employed within a mean-field approximation. The microscopic composition of compact objects, such as Neutron Stars (NSs), remains an open problem. The NSs core is predominantly composed of strongly interacting protons and neutrons at low temperatures and high baryon density. However, as we delve deeper into the inner core, the natural processes during the formation of NSs may give rise to the production of pentaquarks. The existence of light pentaquarks in the heart of NSs has the potential to reduce the maximum mass. This exotic form of matter, known as strange matter, is thought to be abundant in the core of NSs, rendering them among the densest and most exotic objects in the universe. To acquire insights into hyperon-nucleon interactions within the nuclear medium and to figure out the inner structure of NSs, scientists are actively investigating the collective flow of hyper-nuclei in high-energy heavy-ion collisions [38]. The introduction of new particles, such as hyperons, during the star's evolution leads to a temperature drop below nucleon-only stellar matter. The Λ⁰ baryons dominate the heavy baryon composition at all stages of stellar evolution, comprising over 18% of the overall matter content [37]. Noteworthy contributions to hyperon spectroscopy are anticipated from experiments like COMPASS, J-PARC, and the upcoming PANDA experiment [39].

In this context, this work aims to quantitatively investigate, how the nuclear medium modifies the physical properties of the Λ(1405) resonance considering the in-medium dynamics of the Λ(1405). This work operates under the assumption that the Λ(1405) is a pentaquark state rather than a three-quark baryon state. This paper aims to achieve the following objectives: In the next section, we provide a brief overview of the Quantum Chromodynamics Sum Rules (QCDSRs) technique, which is utilized in our computations. Additionally, we discuss the alterations to the properties of Λ(1405) induced by the nuclear medium. Section 3 presents a comprehensive analysis of our results, while Section 4 encapsulates the derived conclusions from these findings.

## 2. The Model

QCDSRs [40–42] stands out as one of the most potent and pragmatic tools utilized to extract the masses and decay constants of hadrons both in a vacuum, nuclear medium and hot medium [43–63]. The QCD vacuum is distinguished by a variety of condensates, including the gluon condensate $\langle G_{\mu\nu}^2 \rangle$, the quark condensate $\langle q\bar{q} \rangle$, the mixed condensate $\langle \bar{q}\sigma Gq \rangle$, and additional ones. It is imperative to incorporate modified condensates when dealing with the nuclear medium.

This study is dedicated to formulating QCDSRs for the Λ(1405) resonance within the framework of the nuclear medium. Our starting point is the covariant time-ordered current-current correlator:

$$\Pi(q) = i \int d^4 x e^{iq \cdot x} \langle \Omega_0 | \mathbf{T}[\mathbf{J}_\Lambda(x) \bar{\mathbf{J}}_\Lambda(0)] | \Omega_0 \rangle. \qquad (2)$$

Here, $q$ denotes the external four-momentum of the $\Lambda(1405)$ state in nuclear matter, $|\Omega_0\rangle$ represents the ground state in the nuclear medium, $\mathbf{T}$ is the time-ordered evolution operator, and $\mathbf{J}_\Lambda(x)$ signifies the hadronic current associated with the $\Lambda(1405)$. Considering the quantum characteristics of spin-1/2 light baryons, we prefer a pentaquark representation composed of $K^- p$ and $\bar{K}^0 n$ for the current associated with the $\Lambda(1405)$ state:

$$\mathbf{J}_\Lambda = \frac{1}{\sqrt{2}}(\mathbf{J}_{K^-}\mathbf{J}_p - \mathbf{J}_{\bar{K}^0}\mathbf{J}_n), \qquad (3)$$

here

$$\mathbf{J}_{K^-} = \bar{u}_l \gamma_5 s_l; \qquad \mathbf{J}_p = \sum_{m=1}^{2} 2\epsilon^{ijk}(u_i^T C A_1^m d_j) A_2^m u_k, \qquad (4)$$

$$\mathbf{J}_{\bar{K}^0} = \bar{d}_l \gamma_5 s_l; \qquad \mathbf{J}_n = \sum_{m=1}^{2} 2\epsilon^{ijk}(d_i^T C A_1^m u_j) A_2^m d_k, \qquad (5)$$

where, $i, j, k$ and $l$ denote the color indices corresponding to the quark fields of $u, d,$ and $s$ quarks, respectively. $C$ denotes the charge conjugation operator, while $T$ symbolizes a transpose in Dirac space. The current is defined with respect to $A_1^m$ and $A_2^m$ as $A_1^1 = I$, $A_1^2 = A_2^1 = \gamma_5$, and $A_2^2 = \beta$. Here, $\beta$ serves as an arbitrary mixing parameter.

The computation of the correlation function in Eq. (2) involves two distinct approaches: the phenomenological (or physical) side, which relies on hadronic characteristics, and the QCD (or theoretical) side, incorporating QCD parameters such as quark masses, in-medium quark, gluon, and quark-gluon mixed condensates. QCDSRs for different physical observables can be derived by equating coefficients of specific structures from both sides. The calculations begin in $x$-space and are subsequently transformed into momentum space. Applying the Borel transformation to both sides of the equation serves to suppress contributions from higher states and the continuum. In the final step of the computation, a continuum subtraction procedure is employed, accompanied by the assumption of quark-hadron duality.

• **Phenomenological approach**: In this perspective, the correlator encapsulates a comprehensive set of in-medium hadronic states, sharing quantum numbers with the corresponding interpolating current. Upon integration over four$-x$, the derived expression is as follows:

$$\Pi^{Phen.}(q) = -\frac{\langle \Omega_0 | \mathbf{J}_\Lambda(x) | \Lambda(q^*, s) \rangle \langle \Lambda(q^*, s) | \bar{\mathbf{J}}_\Lambda(0) | \Omega_0 \rangle}{q^{*2} - m^{*2}} + \cdots. \qquad (6)$$

The notation $|\Lambda(q^*, s)\rangle$ denotes the pentaquark with in-medium four-momentum $q^*$ and spin $s$. The symbol $m^*$ shows the mass in the dense medium. The dots indicate contributions from higher states and the continuum. To carry on, we present the subsequent matrix elements:

$$\langle \Omega_0 | \mathbf{J}_\Lambda(x) | \Lambda(q^*, s) \rangle = \lambda^* u(q, s),$$
$$\langle \Lambda(q^*, s) | \bar{\mathbf{J}}_\Lambda(0) | \Omega_0 \rangle = \bar{\lambda}^* \bar{u}(q, s). \tag{7}$$

Here, $\lambda^*$ denotes the in-medium residue of the $\Lambda(1405)$ resonance, while $u(q, s)$ represents the in-medium Dirac spinor. By substituting Eq. (7) into Eq. (6) and summing over spins, the resulting form of the correlation function's phenomenological side is as follows:

$$
\begin{aligned}
\Pi^{Phen.}(q) &= -\frac{\lambda^{*2}(\not{q}^* + m^*)}{q^{*2} - m^{*2}} + \cdots \\
&= -\frac{\lambda^{*2}}{\not{q}^* - m^*} + \cdots \\
&= -\frac{\lambda^{*2}}{(q_\mu - \Sigma_{v\mu})\gamma^\mu - m^*} + \cdots .
\end{aligned} \tag{8}
$$

The in-medium momentum, denoted as $q_\mu^* = q_\mu - \Sigma_{v\mu}$, and the in-medium mass, $m_\Lambda^* = m_\Lambda + \Sigma^S$, are defined, where $\Sigma_{v\mu}$ and $\Sigma^S$ characterize the in-medium vector and scalar self-energy of the $\Lambda(1405)$ resonance, respectively. These physical quantities can be computed through QCDSRs in the nuclear medium. The contributions of the two independent vectors—particle four-momentum $q_\mu$ and medium four-velocity $u_\mu$—to the vector self-energy are as follows:

$$\Sigma_{v\mu} = \Sigma_v u_\mu + \Sigma_v' q_\mu. \tag{9}$$

In the given framework, $u_\mu$ signifies the four-velocity of the nuclear medium. It is noteworthy that the term $\Sigma_v'$ is neglected owing to its negligible magnitude [62]. Our analysis is performed in the rest frame of the medium, characterized by $u_\mu = (1, 0, 0, 0)$. The correlation function can be decomposed into individual structures, as depicted below:

$$\Pi^{Phen.}(q) = \Pi_{\not{q}}^{Phen.}(q^2, q_0)\not{q} + \Pi_{\not{u}}^{Phen.}(q^2, q_0)\not{u} + \Pi_U^{Phen.}(q^2, q_0)U + \cdots . \tag{10}$$

Here, $U$ symbolizes the unit matrix, and $q_0$ represents the energy of the quasi-particle. The coefficients corresponding to different structures, namely the isoscalar invariant amplitudes $\Pi_i^{Phen.}(q^2, q_0)$, with $i = \not{q}$, $\not{u}$, and $U$ in the aforementioned relation, are derived as:

$$
\begin{aligned}
\Pi_{\not{q}}^{Phen.}(q^2, q_0) &= -\lambda^{*2}\frac{1}{q^2 - \mu^2}, \\
\Pi_{\not{u}}^{Phen.}(q^2, q_0) &= +\lambda^2\frac{\Sigma_v}{q^2 - \mu^2}, \\
\Pi_U^{Phen.}(q^2, q_0) &= -\lambda^{*2}\frac{m^*}{q^2 - \mu^2},
\end{aligned} \tag{11}
$$

where reduced mass can be written as $\mu = \sqrt{m^{*2} - \Sigma_v^2 + 2q_0\Sigma_v}$. Following the application of the Borel transformation concerning $q^2$, the result is then given by:

$$\mathcal{B}\Pi_{\not{p}}^{Phen.}(q^2, q_0) = \lambda^{*2}e^{-\mu^2/M^2},$$
$$\mathcal{B}\Pi_{\not{u}}^{Phen.}(q^2, q_0) = -\lambda^{*2}\Sigma_\upsilon e^{-\mu^2/M^2},$$
$$\mathcal{B}\Pi_{U}^{Phen.}(q^2, q_0) = \lambda^{*2}m^* e^{-\mu^2/M^2}, \tag{12}$$

here the parameter $M^2$ serves as the Borel mass, and its value will be determined in subsequent stages of the analysis.

● **QCD approach**: We inserted the interpolating currents in the correlator and utilized the Wick theorem to contract the quark fields. This process yields the following expression which is the same as the Ref. [13] in connection with the in-medium light quark propagator ($D_q^{ij}$):

$$\Pi^{\text{QCD}}(q) = 4i\epsilon_{abc}\epsilon_{a'b'c'}\int d^4x e^{iq\cdot x}\frac{1}{2}\Bigg\{\text{Tr}[D_u^{d'd}(-x)\gamma_5 D_s^{dd'}(x)\gamma_5]$$

$$\times\Bigg\{-\text{Tr}[D_d^{bb'}(x)\widetilde{D}_u^{aa'}(x)]\gamma_5 D_u^{cc'}(x)\gamma_5 + \gamma_5 D_u^{ca'}(x)\widetilde{D}_d^{bb'}(x)D_u^{ac'}(x)\gamma_5$$

$$-\beta\text{Tr}[D_d^{bb'}(x)\gamma_5\widetilde{D}_u^{aa'}(x)]\gamma_5 D_u^{cc'}(x) + \beta\gamma_5 D_u^{ca'}(x)\gamma_5\widetilde{D}_d^{bb'}(x)D_u^{ac'}(x)$$

$$-\beta\text{Tr}[D_d^{bb'}(x)\widetilde{D}_u^{aa'}(x)\gamma_5]D_u^{cc'}(x)\gamma_5 + \beta D_u^{ca'}(x)\widetilde{D}_d^{bb'}(x)\gamma_5 D_u^{ac'}(x)\gamma_5$$

$$-\beta^2\text{Tr}[D_d^{bb'}(x)\gamma_5\widetilde{D}_u^{aa'}(x)\gamma_5]D_u^{cc'}(x) + \beta^2 D_u^{ca'}(x)\gamma_5\widetilde{D}_d^{bb'}(x)\gamma_5 D_u^{ac'}(x)\Bigg\}$$

$$+\text{Tr}[D_d^{d'd}(-x)\gamma_5 D_s^{dd'}(x)\gamma_5]\Bigg\{-\text{Tr}[D_u^{bb'}(x)\widetilde{D}_d^{aa'}(x)]\gamma_5 D_d^{cc'}(x)\gamma_5$$

$$+\gamma_5 D_d^{ca'}(x)\widetilde{D}_u^{bb'}(x)D_d^{ac'}(x)\gamma_5 - \beta\text{Tr}[D_u^{bb'}(x)\gamma_5\widetilde{D}_d^{aa'}(x)]\gamma_5 D_d^{cc'}(x)$$

$$+\beta\gamma_5 D_d^{ca'}(x)\gamma_5\widetilde{D}_u^{bb'}(x)D_d^{ac'}(x) - \beta\text{Tr}[D_u^{bb'}(x)\widetilde{D}_d^{aa'}(x)\gamma_5]D_d^{cc'}(x)\gamma_5$$

$$+\beta D_d^{ca'}(x)\widetilde{D}_u^{bb'}(x)\gamma_5 D_d^{ac'}(x)\gamma_5 - \beta^2\text{Tr}[D_u^{bb'}(x)\gamma_5\widetilde{D}_d^{aa'}(x)\gamma_5]D_d^{cc'}(x)$$

$$+\beta^2 D_d^{ca'}(x)\gamma_5\widetilde{D}_u^{bb'}(x)\gamma_5 D_d^{ac'}(x)\Bigg\}\Bigg\}. \tag{13}$$

Here, we use the shorthand $\widetilde{D}_q^{ij}(x) = CD_q^{Tij}(x)C$, where $D_q(x)$ stands for the propagators of $u$ or $d$ quarks, $i$ and $j$ denote color indices. We present detailed formulations for the quark propagators within the medium, encompassing crucial elements like in-medium quark, gluon, and mixed condensates. Throughout our calculations, the fixed-point gauge is utilized for the light quark propagator:

$$D_q^{ij}(x) = \frac{i}{2\pi^2}\delta^{ij}\frac{1}{(x^2)^2}\not{x} - \frac{m_q}{4\pi^2}\delta^{ij}\frac{1}{x^2} + \psi_q^i(x)\bar{\psi}_q^j(0)$$

$$-\frac{ig_s}{32\pi^2}\mathcal{F}_{\mu\nu}^A(0)t^{ij,A}\frac{1}{x^2}[\not{x}\sigma^{\mu\nu} + \sigma^{\mu\nu}\not{x}] + \cdots. \tag{14}$$

In the given expression, $\psi_q^i$ and $\bar{\psi}_q^j$ represent Grassmann background quark fields, $\mathcal{F}_{\mu\nu}^A$ signifies classical background gluon fields, and $t^{ij,A} = \frac{\lambda^{ij,A}}{2}$ where $\lambda^{ij,A}$ stands for the standard Gell-Mann matrices. By substituting the light quark propagator from Eq. (14) into the correlation function presented in Eq. (13), then incorporating Grassmann background quark fields and classical background gluon fields, we use terms representing the ground-state matrix elements of related quark and gluon operators from Refs. [56, 63]. We don't give the explicit forms of these expressions for the sake of brevity. On the QCD side, the invariant amplitudes $\Pi_i^{QCD}(q^{*2}, q_0)$ in relation to different structures in Eq. (10) is through the below dispersion integral:

$$\Pi_i^{QCD}(q^2, q_0) = \int_{(3m_u+m_d+m_s)^2}^{s_0^*} \frac{\rho^{QCD}(s, q_0)}{s - q^{*2}} ds. \tag{15}$$

In the above expression, $\rho^{QCD}(s, q_0)$ represents the two-point spectral densities derived from the imaginary parts of the correlation function and $s_0^*$ is the in-medium continuum threshold.

The primary objective on the QCD side is to compute spectral densities using the explicit forms of the in-medium light quark propagator. The calculation involves standard procedures, and although explicit expressions of the spectral functions are not provided here due to their extensive nature, the methodology follows established techniques. As for the QCD side we can write the correlator in terms of separated structures as:

$$\Pi^{QCD}(q) = \boldsymbol{\xi}_{\slashed{q}}^{QCD}(q^2, q_0)\slashed{q} + \boldsymbol{\xi}_{\slashed{u}}^{QCD}(q^2, q_0)\slashed{u} + \boldsymbol{\xi}_U^{QCD}(q^2, q_0)U + \cdots. \tag{16}$$

After rigorous computational analysis, wherein we equate the phenomenological and QCD findings for each structural component, the resulting QCDSRs reveal the mass and residue of the $\Lambda(1405)$ state as follows:

$$\lambda^{*2} e^{-\mu^2/M^2} = \boldsymbol{\xi}_{\slashed{q}}^{QCD}(M^2, s_0^*), \tag{17}$$

$$-\lambda^{*2}\Sigma_v e^{-\mu^2/M^2} = \boldsymbol{\xi}_{\slashed{u}}^{QCD}(M^2, s_0^*), \tag{18}$$

$$\lambda^{*2} m^* e^{-\mu^2/M^2} = \boldsymbol{\xi}_U^{QCD}(M^2, s_0^*), \tag{19}$$

where $\xi_{\slashed{q}, \slashed{u}, U}^{QCD}(M^2, s_0^*)$ are the Borel-transformed invariant functions. The derivation of mass QCDSRs for the $\Lambda(1405)$ state takes various paths from the aforementioned equations. By applying a derivative with respect to $(-1/M^2)$ to both sides of Eq. (17) and eliminating $\lambda^{*2}$ through division, a transformed equation emerges, offering insights into the following:

$$\mu_\Lambda^2 = \frac{d\xi_{\slashed{q}}^{QCD}(M^2, s_0^*)/d\left(-\frac{1}{M^2}\right)}{\xi_{\slashed{q}}^{QCD}(M^2, s_0^*)}. \tag{20}$$

Here, the in-medium mass of the $\Lambda$ state is determined as:

$$m_\Lambda^{*2} = \mu_\Lambda^2 - 2\Sigma_v q_0 + \Sigma_v^2. \tag{21}$$

To determine the in-medium residue, denoted as $\lambda^*$, various possibilities can be used. Using Eq. (17), we can derive the following expression for the residue QCDSRs of the $\Lambda$ state:

$$\lambda^{*2} = \xi_{\slashed{q}}^{QCD}(M^2, s_0^*) e^{\mu^2/M^2}. \tag{22}$$

We employ these extracted QCDSRs to compute the numerical values of both vacuum as the limit $\rho \to 0$ and for the in-medium physical quantities associated with the $\Lambda$ resonance.

## 3. Numerical Results

Here, we undertake a numerical analysis of the obtained QCDSRs previously derived in Eqs. (20-22) to estimate mass and residue of the light spin-1/2 isoscalar $\Lambda(1405)$ state. To accomplish this, we rely on numerical values for various input parameters, including quark masses, as well as in-medium and vacuum condensates. These condensates include quark, gluon, and mixed condensates of different dimensions, and the specific values are detailed in Table (1). To determine the reliable range of the auxiliary parameter $\beta$, we plot the QCD side of the result obtained from the $\not{q}$ structure for the $\Lambda(1405)$ state. As a parameter, we utilize $x = \cos\theta$, where $\theta = \arctan\beta$ and $\beta = -0.82$. Remarkably, the medium significantly expands the reliable ranges of $\beta$, marking a key outcome of this study. The QCDSRs for the physical quantities outlined in Eq. (20-22) consist three auxiliary parameters: the Borel mass parameter $M^2$, the in-medium continuum threshold $s_0^*$, and the mixing parameter $\beta$ that influences the symmetric and anti-symmetric spin-1/2 currents. We aim to identify their operational ranges in line with established methodological guidelines, ensuring that the physical quantities exhibit weak dependence on these parameters within these ranges.

| Parameter | Numeric | Unit |
|---|---|---|
| $m_u$ | $2.16^{+0.49}_{-0.26}$ | MeV |
| $m_d$ | $4.67^{+0.48}_{-0.17}$ | MeV |
| $m_s$ | $93.4^{+8.6}_{-3.4}$ | MeV |
| $m_\Lambda$ | $1405^{+1.3}_{-0.9}$ | MeV |
| $m_q$ | $0.0034$ | GeV |
| $\sigma_N$ | $0.045$ | GeV |
| $\sigma_{N_0}$ | $0.035$ | GeV |
| $m_0^2$ | $0.8$ | GeV$^2$ |
| $\rho^{sat}$ | $(0.11)^3$ | GeV$^3$ |
| $\langle q^\dagger q \rangle_\rho$ | $\frac{3}{2}\rho$ | GeV$^3$ |
| $\langle s^\dagger s \rangle_\rho$ | $0$ | GeV$^3$ |
| $\langle \bar{q}q \rangle_0$ | $(-0.241)^3$ | GeV$^3$ |
| $\langle \bar{s}s \rangle_0$ | $0.8\langle \bar{q}q \rangle_0$ | GeV$^3$ |
| $\langle \bar{q}q \rangle_\rho$ | $\langle \bar{q}q \rangle_0 + \frac{\sigma_N}{2m_q}\rho$ | GeV$^3$ |
| $y$ | $1 - \frac{\sigma_{N_0}}{\sigma_N}$ | – |
| $\langle \bar{s}s \rangle_\rho$ | $\langle \bar{s}s \rangle_0 + y\frac{\sigma_N}{2m_q}\rho$ | GeV$^3$ |
| $\langle q^\dagger iD_0 q \rangle_\rho$ | $0.18\rho$ | GeV$^4$ |
| $\langle s^\dagger iD_0 s \rangle_\rho$ | $\frac{m_s\langle \bar{s}s \rangle_\rho}{4} + 0.02 \text{ GeV}\rho$ | GeV$^4$ |
| $\left\langle \frac{\alpha_s}{\pi}G^2 \right\rangle_0$ | $(0.33 \pm 0.04)^4$ | GeV$^4$ |
| $\left\langle \frac{\alpha_s}{\pi}G^2 \right\rangle_\rho$ | $\left\langle \frac{\alpha_s}{\pi}G^2 \right\rangle_0 - 0.65 \text{ GeV}\rho$ | GeV$^4$ |
| $\langle \bar{q}g_s\sigma Gq \rangle_0$ | $m_0^2\langle \bar{q}q \rangle_0$ | GeV$^5$ |
| $\langle \bar{q}g_s\sigma Gq \rangle_\rho$ | $\langle \bar{q}g_s\sigma Gq \rangle_0 + 3\text{GeV}^2\rho$ | GeV$^5$ |
| $\langle \bar{s}g_s\sigma Gs \rangle_0$ | $m_0^2\langle \bar{s}s \rangle_0$ | GeV$^5$ |
| $\langle \bar{s}g_s\sigma Gs \rangle_\rho$ | $\langle \bar{s}g_s\sigma Gs \rangle_0 + 3y\rho$ | GeV$^5$ |
| $\langle q^\dagger iD_0 iD_0 q \rangle_\rho$ | $0.031 \text{ GeV}^4\rho - \frac{1}{12}\langle \bar{q}g_s\sigma Gq \rangle_\rho$ | GeV$^5$ |
| $\langle s^\dagger iD_0 iD_0 s \rangle_\rho$ | $0.031y \text{ GeV}^2\rho - \frac{1}{12}\langle \bar{s}g_s\sigma Gs \rangle_\rho$ | GeV$^5$ |

**Table 1.** Parameters employed as inputs in the calculations [5, 63–68].

For this purpose, we adhere to standard prescriptions of the method, demanding both pole dominance and the convergence of operator product expansion series. In technical terms, determining the upper limit of the Borel mass parameter involves ensuring that the pole contribution exceeds the contributions from higher states and the continuum. The lower boundary of $M^2$ is determined by requiring that the perturbative component surpasses the non-perturbative contributions and ensures convergence of the series containing non-perturbative operators. In this context, we set the Borel mass to guarantee that the pole contribution consistently constitutes a range between 80% and 50% of the total contribution. The choice of the continuum threshold is not arbitrary but is contingent on the energies of the first excited states in the relevant channels.

The considered resonance, when subjected to these criteria, meets all the necessary conditions. The mass and residue in the limit as $\rho \to 0$, denoted as $m_\Lambda$, $\lambda_\Lambda$ exhibits notable stability, as depicted in Fig. (1), and illustrates that the mass and residue variations concerning the continuum threshold are minimal within the selected window. The ranges of Borel mass parameter $M^2$ and the in-medium continuum threshold $s_0^*$ satisfy all the procedural requirements as:

$$M^2 \in [1.5 - 2.0] \text{ GeV}^2, \quad s_0^* \in [2.3 - 2.9] \text{ GeV}^2. \tag{23}$$

We also illustrate the ratio of the in-medium mass to the vacuum mass $m_\Lambda^*/m_\Lambda$ and

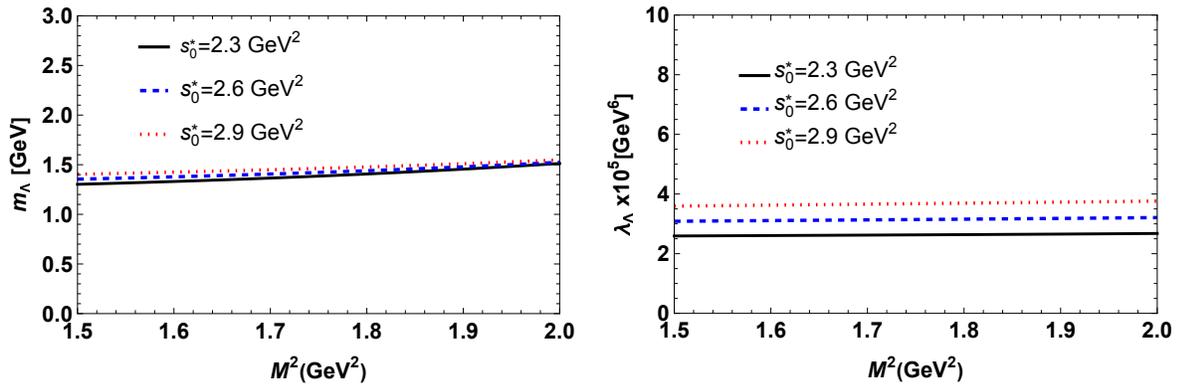

**Figure 1.** (a) Change of the $\Lambda(1405)$ vacuum mass in terms of $M^2$ Borel mass parameter. (b) The same graph for the residue.

$\lambda_\Lambda^*/\lambda_\Lambda$ in Fig. (2) below with respect to $M^2$ for the fully light $\Lambda$ state at different values of the continuum threshold and the saturation nuclear matter density. The depicted figure demonstrates the strong stability of $m_\Lambda^*/m_\Lambda$ across the chosen $M^2$ windows for all constituents. Notably, the $\Lambda(1405)$ experiences a considerable effect from the medium, decreasing to approximately 80% and 62% of its vacuum value at saturation nuclear matter density for both the mass and residue, respectively. It is crucial to emphasize that the vacuum masses are extrapolated from in-medium calculations in the limit as $\rho$ tends to zero.

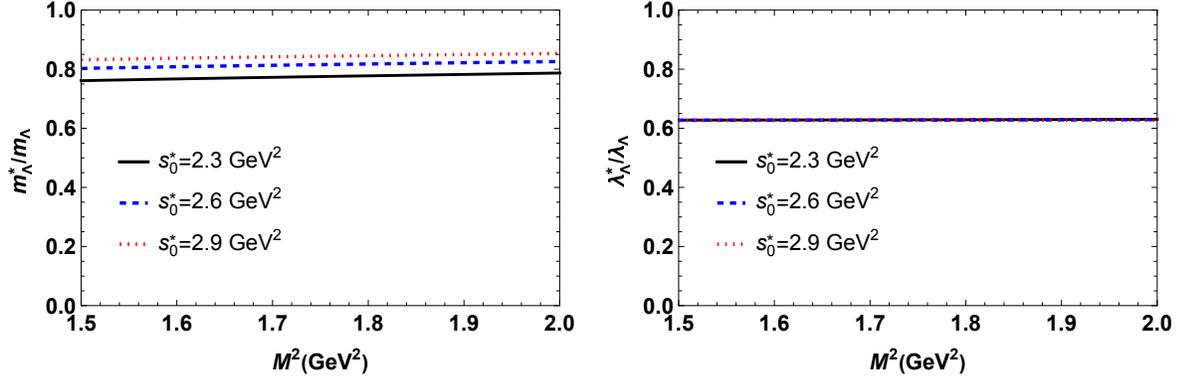

**Figure 2.** (**a**) The ratio of the in-medium mass to the vacuum mass for the $\Lambda(1405)$ as a function of $M^2$ Borel mass parameter at different values of the continuum threshold and saturation density. (**b**) The same plot for residue.

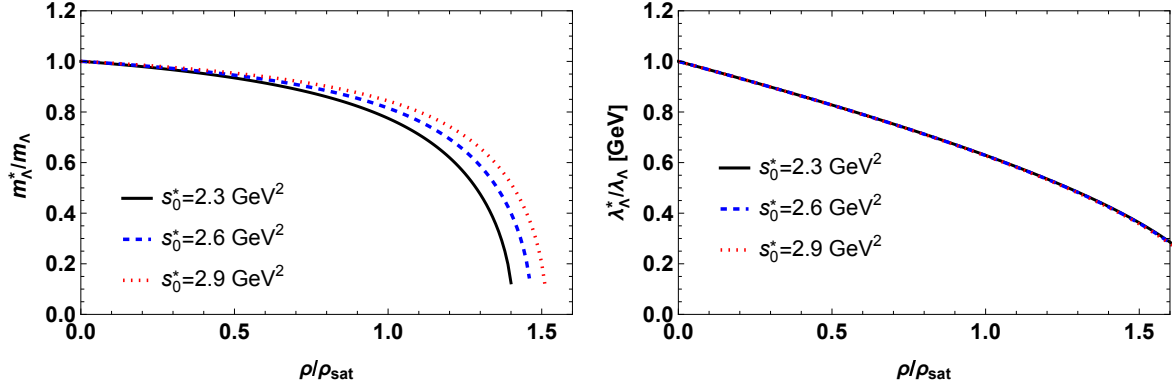

**Figure 3.** (**a**) The ratio of the in-medium mass to the vacuum mass for the $\Lambda(1405)$ as a function of $\rho/\rho_{sat}$ at different values of the continuum threshold and the mean value of Borel mass parameter. (**b**) In-medium residue variation depicted in the accompanying graph.

The main objective of this study is to examine how the mass and residue of the states under consideration behaves with the density of the medium. In Fig. (3), we illustrate the ratio of nuclear medium mass to vacuum mass $m_\Lambda^*/m_\Lambda$ and $\lambda_\Lambda^*/\lambda_\Lambda$ versus $\rho/\rho_{sat}$ for $\Lambda(1405)$ at different values of the continuum threshold, mean value of Borel mass parameter, and within the reliable regions of the mixing parameter $\beta$. Meanwhile we explore the mass and residue behavior within the range $\rho \in [0, 1.6]\rho_{sat}$ as it is seen in Fig. (3).

The pentaquark configuration of the $\Lambda(1405)$ resonance, containing one strange, three up, and one down quark, experiences a substantial mass reduction, reaching nearly 10% of its vacuum value at $\rho/\rho_{sat} = 1.5$. As illustrated in Fig. (3), at $\rho/\rho_{sat} = 1$, the in-medium mass to vacuum mass ratio stands at roughly 0.80 for this resonance. For the residue, these values correspond to $\sim 36\%$ and $\sim 0.62$.

The saturation nuclear matter density holds particular significance in our study. In Table (2), we present numerical values for the modified masses of the $\Lambda(1405)$ state at $\rho = \rho_{sat}$, alongside the vacuum masses of the fully light pentaquark candidate $\Lambda(1405)$ in the limit as $\rho \to 0$. The numerical results exhibit uncertainties attributed to errors in input parameters and uncertainties in determining the optimal working windows for auxiliary parameters.

Upon comparing vacuum mass with mass at the saturation point, it becomes evident that the $\Lambda(1405)$ state responds to its environment in a dense medium. In summary, we compare the mass values of the fully light $\Lambda(1405)$ state with spin $-1/2$ using QCDSRs in the $\rho \to 0$ limit. Table (2) and (3) serve as the focal point for this comparison, encompassing theoretical predictions and existing experimental data within the $\Lambda(1405)$ channel. A careful examination of the table shows that results obtained through various methods [13, 17] are generally consistent, falling within mutually acceptable margins of error.

| State | $m(\rho = 0)$ | $m^*(\rho = \rho_{sat})$ |
|---|---|---|
| $\Lambda(1405)$ | $1.408^{-0.126}_{+0.094}$ | $1.141^{-0.122}_{+0.120}$ |

**Table 2.** The mass values of $\Lambda(1405)$ at vacuum and the mean modified mass at the saturation nuclear matter density in GeV.

| State | $\lambda(\rho = 0)$ | $\lambda^*(\rho = \rho_{sat})$ |
|---|---|---|
| $\Lambda(1405)$ | $3.152^{-0.559}_{+0.610} \times 10^{-5}$ | $1.982^{-0.312}_{+0.364} \times 10^{-5}$ |

**Table 3.** The residue characteristic of $\Lambda(1405)$ under vacuum condition, alongside the mean modified residue at saturation nuclear matter density in unit of GeV$^6$.

Our estimate for the mass of $\Lambda(1405)$, determined to be $1408^{-126}_{+94}$ MeV, harmoniously agrees with the PDG's mean value of $1405.1^{+1.3}_{-1.0}$ MeV. This congruence in predictions, not only for the $\Lambda(1405)$ but also for other related members, coupled with insights from various theoretical models, holds promise in illuminating future experiments. These experiments are poised to delve into the realm of fully light pentaquarks, offering the prospect of measuring their properties meticulously.

## 4. Conclusion

In recent years, there has been a profound exploration into the properties of hadrons, specifically mesons with strangeness, within dense matter. This investigation is closely intertwined with the examination of exotic particles and the analysis of heavy-ion collisions. Particularly, the elusive nature of strange matter is believed to be abundant in the cores of neutron stars, marking them as among the densest and most exotic entities in the universe. While the notion of physically journeying to neutron stars for the study of this exotic matter remains firmly in the realm of science fiction, valuable insights into these celestial bodies can be gleaned through particle collisions conducted in heavy-ion collisions. This line of inquiry holds the potential not only to constrain existing theories but also to offer variable inputs for calculations, providing valuable information about the phase transition, structure, and composition of compact stars. The pursuit of understanding less-known resonances and confirming their existence plays a crucial role in establishing the multiple structures of hadrons.

Addressing the nature of $\Lambda(1405)$ within conventional three-quark models poses challenges. Our investigation focuses on the mass and residue of $\Lambda(1405)$ with the assumption of a pentaquark state in a cold nuclear medium. Employing in-medium QCDSRs up to dimension six, our study specifically explores the properties of $\Lambda(1405)$ considering it as a molecule and nucleon in a dense medium. The nuclear medium effect emerges as a significant factor, substantially influencing the in-medium characteristics of $\Lambda(1405)$, notably its mass and residue. According to our results we find the mass $1.141^{-0.122}_{+0.120}$ GeV and residue $1.982^{-0.312}_{+0.364} \times 10^{-5}$ GeV$^6$ at saturation density and $1.408^{-0.126}_{+0.094}$ GeV, $3.152^{-0.559}_{+0.610} \times 10^{-5}$ GeV$^6$ at vacuum. In the presence of nuclear matter, there is an approximately 20% change in the mass value of $\Lambda(1405)$, along with a 38% alteration in the residue value. Our results, in the limit as $\rho_{sat} \to 0$, are consistent with experimental data [5] and support the meson-baryon molecular structure proposed for $\Lambda(1405)$ in Ref. [13]. The combined findings from experiments, lattice QCD simulations, and theoretical studies on the $\Lambda(1405)$ resonance are set to transform our understanding of hadron and nuclear physics.

## 5. Acknowledgement

We are grateful to K. Azizi for his valuable discussions.